\documentclass[twocolumn,prb,showpacs,preprintnumbers,amsmath,amssymb,superscriptaddress]{revtex4}

\usepackage{graphicx}
\usepackage{dcolumn}
\usepackage{bm}

\begin{document}


\title{Coherent versus Incoherent Light Scattering from a Quantum Dot}

\author{K. Konthasinghe}%
\affiliation{%
Dept. of Physics, University of South Florida, Tampa, FL 33620}

\author{J. Walker}
\affiliation{%
Dept. of Physics, University of South Florida, Tampa, FL 33620}

\author{M. Peiris}%
\affiliation{%
Dept. of Physics, University of South Florida, Tampa, FL 33620}

\author{C. K. Shih}%
\affiliation{%
University of Texas at Austin, Austin, TX 78712}

\author{Y. Yu}%
\affiliation{%
Institute of Semiconductors, Chinese Academy of Sciences, Being, PR China}

\author{M. F. Li}%
\affiliation{%
Institute of Semiconductors, Chinese Academy of Sciences, Being, PR China}

\author{J. F. He}%
\affiliation{%
Institute of Semiconductors, Chinese Academy of Sciences, Being, PR China}

\author{L. J. Wang}%
\affiliation{%
Institute of Semiconductors, Chinese Academy of Sciences, Being, PR China}

\author{H. Q. Ni}%
\affiliation{%
Institute of Semiconductors, Chinese Academy of Sciences, Being, PR China}

\author{Z. C. Niu}%
\affiliation{%
Institute of Semiconductors, Chinese Academy of Sciences, Being, PR China}

\author{A. Muller}%
\email{mullera@usf.edu}
\affiliation{%
Dept. of Physics, University of South Florida, Tampa, FL 33620}

\date{\today}

\begin{abstract} 
We analyze the light scattered by a single InAs quantum dot interacting with a resonant continuous-wave laser. High resolution spectra reveal clear distinctions between coherent and incoherent scattering, with the laser intensity spanning over four orders of magnitude. We find that the fraction of coherently scattered photons can approach unity under sufficiently weak or detuned excitation, ruling out pure dephasing as a relevant decoherence mechanism. We show how spectral diffusion shapes spectra, correlation functions, and phase-coherence, concealing the ideal radiatively-broadened two-level system described by Mollow.
\end{abstract}

\maketitle

\pagebreak

Like an isolated atom or ion, a semiconductor quantum dot (QD) ``artificial atom'' scatters monochromatic laser light incident upon it. For an ideal two-level system broadened by radiative decay at a rate $\kappa$, the spectral and temporal properties of the scattered photons are determined solely by the laser detuning from resonance, $\Delta\omega$, and by the Rabi frequency, $\Omega$. In seminal work, Mollow showed that when $\Omega\ll\kappa$, most of the light is scattered {\it coherently}, i.e. elastically, whereas it is otherwise dominated by resonance fluorescence \cite{mollow1969psl}. Moreover, when the scattered light originates from a single two-level system it exhibits photon anti-bunching, {\it for any value of} $\Omega$ {\it and} $\Delta\omega$ \cite{cohentannoudji}.

Probing and controlling quantum phase coherence is at the core of quantum information science, and semiconductor QDs are well suited for investigating quantum coherence in solids at optical frequencies \cite{bayer2000hse}. In this context, resonant light scattering in QDs has been of special interest, and has emerged as a promising resource for the generation of highly ideal single photon states \cite{kiraz2004qds}. Milestone demonstrations include oscillatory field correlations \cite{muller2007rfc}, Mollow triplets \cite{flagg2009, vamivakas2009, ates2009}, photon anti-bunching \cite{flagg2009} and cascaded photon emission \cite{ulhaq2011hsp}. Nevertheless, the theoretical modelling of these experimental observations has necessitated the inclusion of a phenomenological dephasing time, $T_2$, in general smaller than 2$T_1$, where $T_1=1/\kappa$ is the radiative lifetime \cite{muller2007rfc, flagg2009, vamivakas2009, ates2009, ulhaq2011hsp}.  Moreover, the {\it coherent} scattering---as opposed to incoherent scattering that causes resonance fluorescence---has remained largely unexplored. A more detailed experimental investigation clarifying the role of decoherence in the context of resonant light scattering is all the more timely given recent proposals of using coherent scattering for the generation of indistinguishable photons \cite{nguyen2011ucs, matthiesen2011snl}: theory predicts that dephasing reduces the fraction of coherently scattered photons from unity, regardless of $\Omega$ and $\Delta\omega$. Here we combine high resolution spectroscopy (35 MHz), photon correlations and phase-coherence measurements, to obtain clear distinctions between coherent and incoherent scattering. We show that the previously observed non-ideal linewidth and correlation functions are due to the fluctuation of the QD resonance frequency on a time scale longer than $T_1$, resulting in an apparent broadening of the lineshape. One the other hand, at the relevant time scale of quantum evolution, the system is close to an ideal two-level system with negligible pure dephasing, as described by Mollow \cite{mollow1969psl}, and as expected from four-wave mixing measurements on QD ensembles \cite{borri2001udt,langbein2004rld}. We extend Mollow's theory to include spectral diffusion in the form of inhomogeneous broadening and provide a complete picture of resonant light scattering in semiconductor QDs.

We probe QDs grown by molecular beam epitaxy at the center of a planar optical microcavity. QDs of this type have been investigated extensively due to their atom-like spectra \cite{bayer2000hse} and long dephasing times \cite{borri2001udt}. The QD sample was grown using a solid source VEECO Gen-II molecular beam epitaxy (MBE) system on a semi-insulating GaAs (100) substrate with a 12 pairs top and 20 pairs bottom Al$_{0.9}$Ga$_{0.1}$As/GaAs distributed Bragg reflector. The substrate rotation was stopped during MBE growth of InAs layers to obtain a QD density varying uniformly from 10$^9$/cm$^2$ to 10$^8$/cm$^2$ \cite{huang2007}. The dominant vertical cavity mode is centered around $\lambda$$\approx$925 nm. The sample was maintained at a temperature of 3.8 K in a closed-cycle cryostat and the QD emission was collected by an {\it in situ} high numerical aperture aspheric lens [Fig. 1(a)]. To excite a QD resonantly and discriminate the resonant scattering signal from stray laser light, an orthogonal excitation/detection geometry was used \cite{muller2007rfc}. The detected light was coupled into a single mode fiber without prior filtering or cross-polarized excitation/detection. For coarse spectral analysis of the emitted light a grating spectrometer with a cooled charge-coupled device camera was employed. For high resolution spectral measurements we used a scanning Fabry-Perot interferometer with a free spectral range of 4.2 GHz and finesse of 120 in conjunction with a single photon counting detector. The excitation source was a diode laser continuously tunable over a range of several GHz. We have probed a number of single QDs and found very similar behavior, although the data presented here is from one specific QD.  

\begin{figure}[t!]
\includegraphics[width=3.3in]{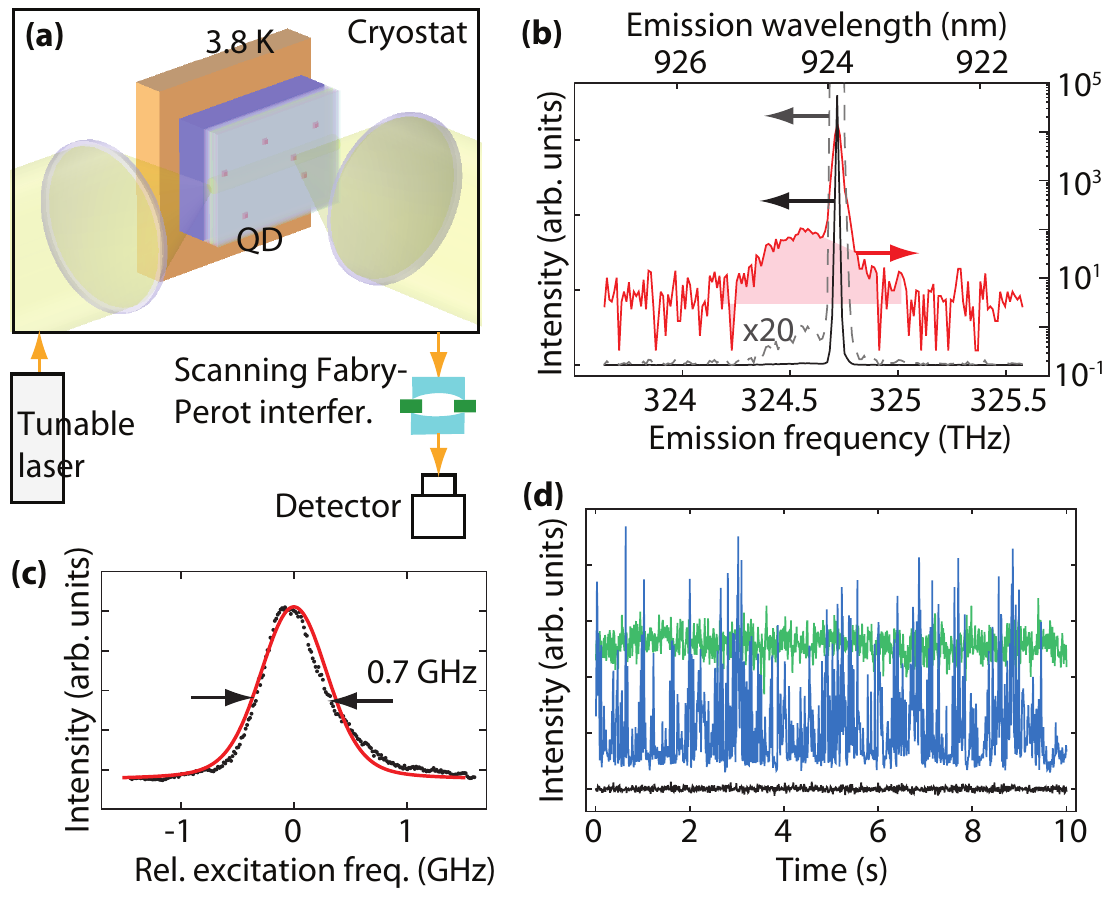}
\caption{\label{fig1} (Color online) (a) Schematic of experimental setup. (b) QD emission spectrum, recorded with a grating spectrometer under resonant laser excitation as is (black solid trace), 20$\times$ magnified (dashed gray trace) and on a logarithmic scale (solid red trace) to visualize the phonon broadband. (c) QD excitation spectrum. (d) Spectrally integrated intensity of scattered light as a function of time showing flickering (dark blue trace) that is inhibited when an additional weak auxiliary laser is added (light green trace). The black trace was recorded with the auxiliary laser only.}
\end{figure}

Figure 1(b) shows the power spectrum of this QD under resonant laser excitation, recorded with a spectrometer with $\approx$10 GHz resolution. Although the main emission line is not resolved in this measurement, a broadband emission around it can be clearly identified. As is well-known from experimental photoluminescence (PL) measurements \cite{besombes2001apb}, four-wave mixing studies \cite{borri2001udt,langbein2004rld}, and from theoretical calculations of resonance fluorescence spectra \cite{ahn2005rfs}, this broad emission originates from fast scattering processes with acoustic phonons. Although at 3.8 K [Fig. 1(b)] this phonon scattering ``pedestal'' is highly asymmetric it becomes more symmetric and prominent with increasing temperature \cite{besombes2001apb, borri2001udt, ahn2005rfs}. In Fig. 1(b) about 5\% of light is emitted into this broad band, thus at liquid He temperature as much as $\approx$95\% of light may be scattered coherently, irrespective of $\Omega$ and $\Delta\omega$.

\begin{figure*}[t!]
\includegraphics[width=5.94in]{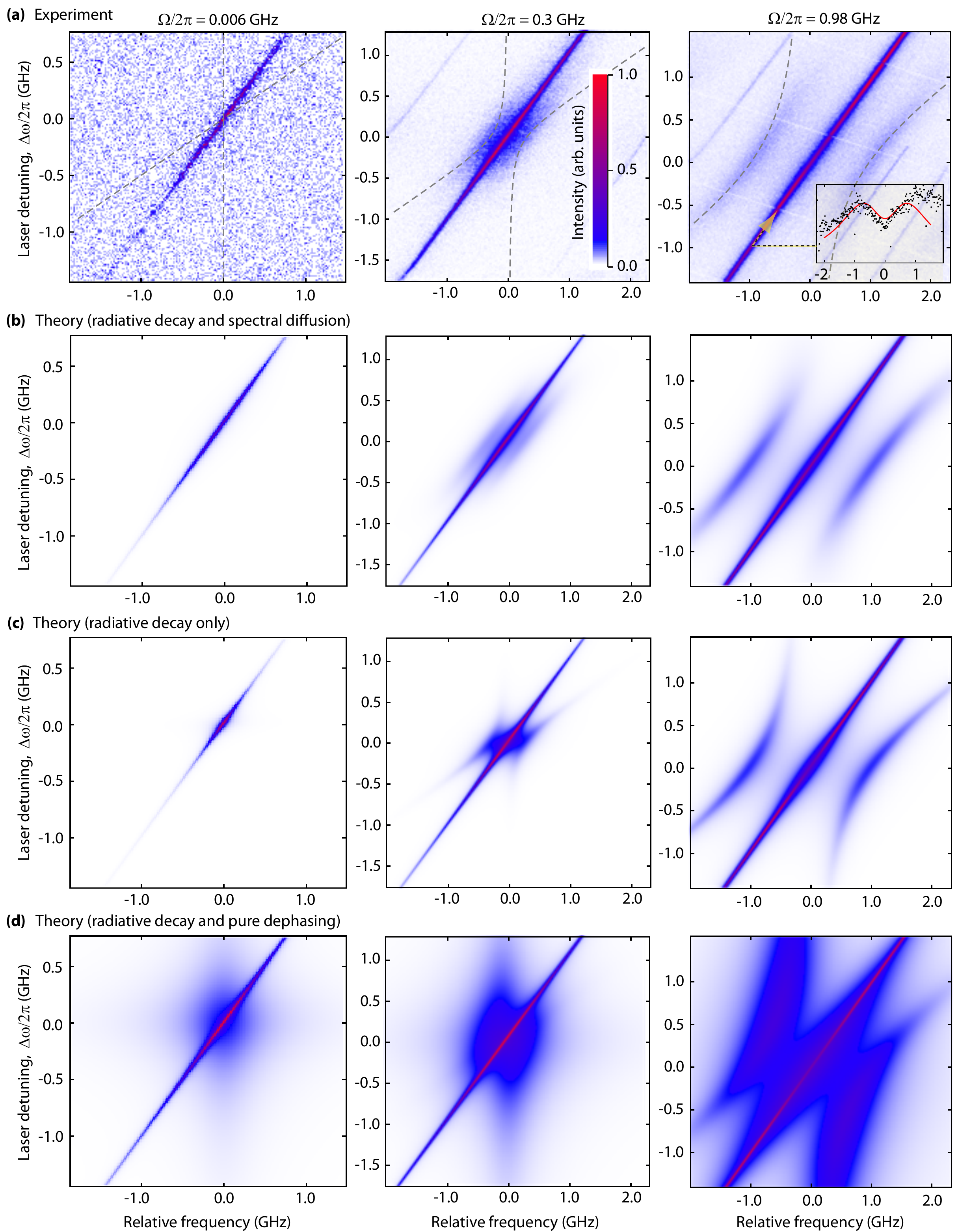}
\caption{\label{fig2} (Color online) (a) Maps of scattered light intensity as a function of detection frequency (abscissas) and excitation frequency (ordinates) relative to the QD transition frequency, for three different values of the Rabi frequency. Each panel was recorded in 200 s. At $\Omega/2\pi$=0.13 GHz the detector count rate was 3$\times10^5$ s$^{-1}$ at the input of the Fabry-Perot interferometer. The dashed lines indicate the location of the eigenfrequencies of the coupled laser/QD system. The inset shows the line section $\nu=\omega$ together with the corresponding theoretical curve, obtained from Eq. (2). The faint diagonal lines parallel to the $\nu=\omega$ section are due to the residual transmission at high-order modes of the Fabry Perot interferometer and satellite modes of the laser. (b) Theoretical maps [Eq. (2)] corresponding to a radiatively-broadened two-level system subject to spectral diffusion. (c) Theoretical maps [Eq. (1)] representing an ideal, radiatively-broadened two-level system. (d) Theoretical maps [Eq. (A11)] representing a two-level system subject to radiative decay, and to pure depahsing at a rate $\gamma$. $\gamma$ is chosen so as to obtain the excitation linewidth of Fig. 1(c).}
\end{figure*}

Figure 1(c) shows the QD excitation spectrum at an excitation intensity below the saturation intensity, obtained by scanning the laser across the QD resonance frequency while collecting the scattered light. The full width at half maximum (FWHM) of the resonance is about 0.7 GHz. Although in four-wave mixing studies the zero-phonon linewidth has been shown to be limited by radiative decay to about 170 MHz \cite{borri2001udt, langbein2004rld}, PL linewidths and/or resonantly measured single QD excitation linewidths are usually found to lie in the range of 500 MHz to several GHz \cite{muller2007rfc,nguyen2011ogr,flagg2009, vamivakas2009, ates2009, nguyen2011ucs, matthiesen2011snl,ulrich2011dms, xu2007cos, hogele2004vco, ulhaq2011hsp}. The additional ``apparent'' broadening is due to spectral diffusion, a process by which the QD transition frequency is randomly shifted during the measurement \cite{hogele2004vco}. This shift is thought to originate from a fluctuating charge environment of the QD that occurs on a time scale that is long compared to the radiative decay process. In PL measurements spectral diffusion is manifested in the form of long-time photon correlations and blinking \cite{santori2004scp}. In resonant scattering experiments the existence of charge fluctuations is evidenced by flickering \cite{latta2009crl} and by the fact that many resonances identified in PL as neutral exciton transitions generate little or no resonance fluorescence unless a weak background lighting is present \cite{nguyen2011ogr}. It has been proposed that a Coulomb blockade effect is at the origin of this resonant emission quenching \cite{nguyen2011ogr}. Flickering of the resonant scattering signal is directly observed when no external non-resonant lighting is present, as seen in Fig. 1(d), consistent with spectral diffusion \cite{neuhauser2000cbf}. The flickering can be inhibited on the measurement timescale ($\sim$10 ms) by an auxiliary non-resonant light source (here a laser at $\lambda$=660 nm) that is so weak that by itself generates negligible PL [Fig. 1(d)]. We speculate that the effect of this auxiliary laser is to change the timescale of the flickering by neutralization of surrounding charges. Spectral diffusion is also commonly observed for QDs in nanostructures, where it is associated with the proximity of etched surfaces \cite{majumdar2011epc}.

In order to distinguish experimentally between coherent and incoherent scattering it is necessary to analyze the scattered light spectrally with a resolution better than $\kappa$.
For a two-level system with natural resonance frequency $\omega_0$, exposed to a monochromatic field of frequency $\omega$, the power spectrum of the scattered light takes the analytic form, given by Mollow \cite{mollow1969psl},
\begin{equation}
\tilde{g}(\nu,\omega,\omega_0)=2\pi|\alpha_{\infty}|^2\delta(\nu-\omega)
+\bar{n}_{\infty}\kappa\Omega^2\frac{(\nu-\omega)^2+\Omega^2/2+\kappa^2}{|f(i(\nu-\omega))|^2}
\end{equation}
where $\nu$ is the emission frequency, and the only source of broadening is a decay of the upper state to the lower state at a rate $\kappa$ due to spontaneous emission. The first term corresponds to photons scattered coherently while the second term describes the resonance fluorescence. The laser detuning $\Delta\omega=\omega-\omega_0$ enters through the polynomial function $f$, the steady-state population inversion $\bar{n}_{\infty}$, and the steady-state quantum mechanical expectation value of the two-level coherence, $\alpha_\infty$, given in \cite{mollow1969psl} and in Appendix A. Note that the function $\tilde{g}(\nu,\omega,\omega_0)$ is always symmetric around $\nu=\omega$ (not $\nu=\omega_0$), unlike spectra of dressed states populated non-resonantly \cite{jundt2008ode,muller2008esd}.

Figure 2(a) shows maps of the scattered light as a function of emission and excitation frequency, for various values of the Rabi frequency. The latter was varied by varying the laser intensity. 
As expected from Eq. (1), when $\Omega\ll\kappa$ [leftmost map in Fig. 2(a)], the emission resonance linewidth is much less than $\kappa$. It is limited here by the resolution of our scanning Fabry-Perot interferometer (35 MHz) but is expected to be as narrow as the laser linewidth ($\sim$1 MHz). With increasing $\Omega$ the resonance fluorescence eventually dominates in the form of the Mollow triplet, composed of three peaks with a FWHM that is of the order of $\kappa$ [middle and rightmost map in Fig. 2(a)].

In order to identify the dominant effects on the QD two-level system due to its solid-state environment, we make a side-by-side comparison of our experimental data of Fig. 2(a) with spectral maps obtained from three different theoretical models, displayed in Fig. 2(b-d). In all cases, we assume a radiative decay, $\kappa/2\pi$=180 MHz, which is consistent with typical radiative lifetimes [$\tau$=1/(2$\pi\times$180 MHz)=0.9 ns] for InAs QDs. We include spectral diffusion by integrating Eq. (1) over all possible (random) detunings to obtain
\begin{equation}
I(\nu,\omega,\omega_0)\propto \int\tilde{g}(\nu,\omega,\omega_0') e^{-(\omega_0'-\omega_0)^2/2\sigma^2} d\omega_0'
\end{equation}
where we assume a Gaussian distribution of QD resonance frequencies due to spectral diffusion with FWHM $s\approx2.355\sigma$. While spectral diffusion may manifest at different timescales \cite{favero2007tdz}, we assume here that this timescale is long compared to the radiative decay time. We also neglect any influence of dark excitonic states \cite{bayer2000ssd}. To account for finite apparatus resolution we replace the delta function in Eq. (1) by a normalized Lorentzian with FWHM equal to 35 MHz. Fig. 3(b) shows spectral maps obtained using Eq. (2) with $s/2\pi$=0.7 GHz, chosen to coincide with the measured excitation linewidth of Fig. 1(c) which is also typical for InAs QDs \cite{muller2007rfc,nguyen2011ogr,flagg2009, vamivakas2009, ates2009, nguyen2011ucs, matthiesen2011snl,ulrich2011dms, xu2007cos, hogele2004vco, ulhaq2011hsp, latta2009crl}. In comparison, Fig. 2(c) shows spectral maps obtained using Eq. (1) clearly not agreeing as well with the experimental data of Fig. 2(a) and highlighting the subtle effects of the spectral diffusion process. In Fig. 2(d) we further show the case when pure dephasing is present instead of spectral diffusion, as was assumed in previous studies \cite{muller2007rfc,flagg2009}. The most general theoretical expressions for the power spectrum for the latter case are derived in Appendix A. The pure dephasing rate, $\gamma$, was chosen to yield the observed excitation linewidth [Fig. 1(c)]. As is evident from this comparison, pure dephasing does not play a major role at the measurement temperature since it would give rise to rather different observations. In particular, the absence of broad emission superimposed to the leftmost spectral map in Fig. 2(a) is an unequivocal indication that dephasing is not a relevant broadening process here.

\begin{figure}[b!]
\includegraphics[width=3.5in]{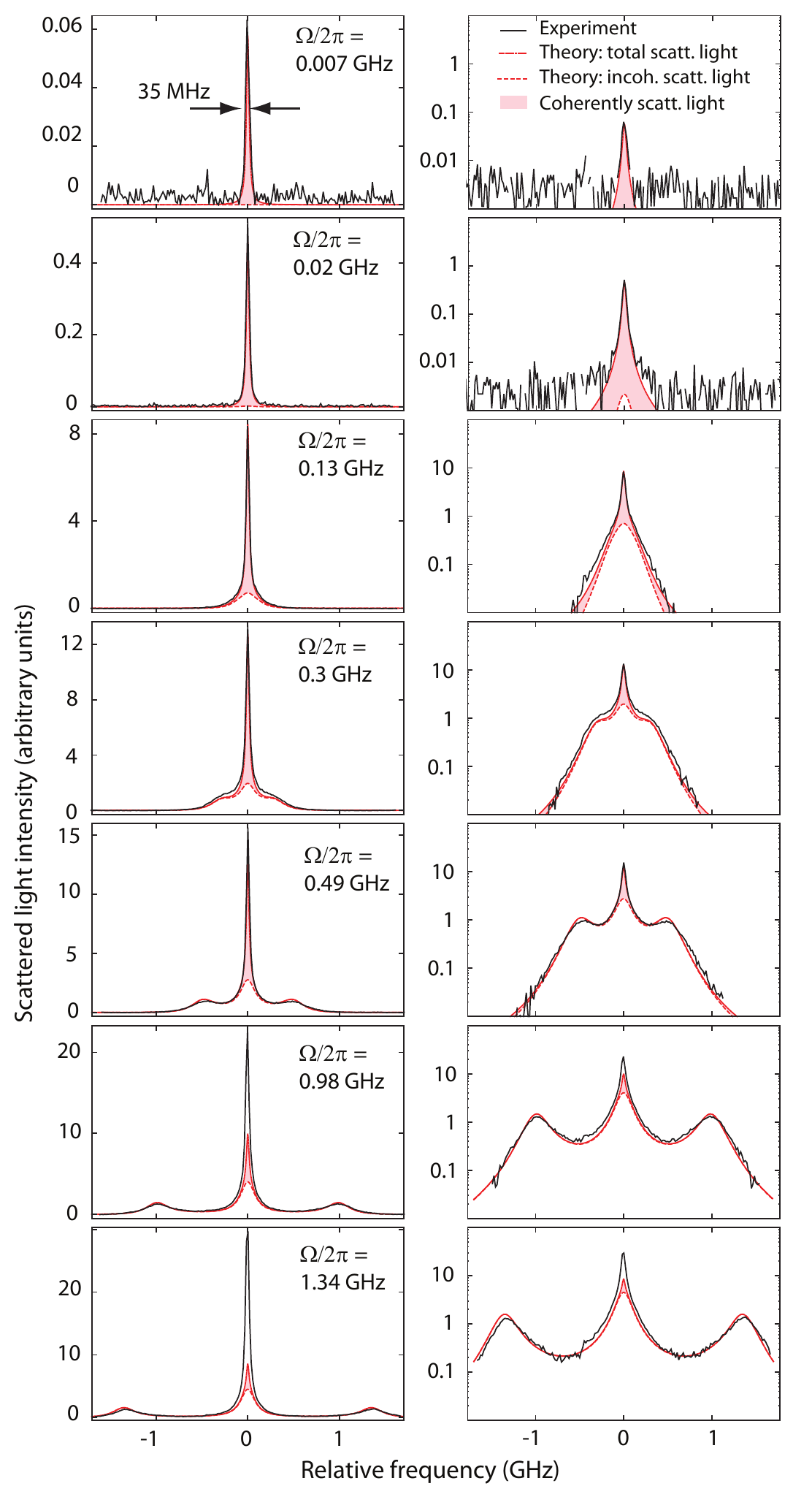}
\caption{\label{fig3} (Color online) Power spectrum of light scattered by the QD at exact resonance ($\Delta\omega=0$) represented on a linear (left) and logarithmic (right) ordinate scale, for a range of Rabi frequencies. Each spectrum was recorded in 60 s. The theoretical curve (red dashed line) was obtained by numerical evaluation of Eq. (2). The light red shaded area corresponds to coherently scattered light which dominates whenever $\Omega\lesssim\kappa$. If pure dephasing had a significant role in the scattering process, a broader feature would always be visible, even when $\Omega\ll\kappa$.}
\end{figure}

For a quantitative analysis we recorded a series of spectra at exact resonance ($\Delta\omega=0$) with sufficiently high signal to noise ratio for precise comparison with theory (Fig. 3). The same information is displayed on both linear (left) and logarithmic (right) ordinate scales. The theoretical traces were obtained using Eq. (2) with the same parameters as in Fig. 2(b), and only a common scale factor was permitted for all traces. Short-dashed (long-dashed) red lines correspond to the total (incoherent) scattered light intensity and the light-red shaded area indicates the coherently scattered light.
As is seen in the figure, there is excellent agreement between the experimental data and the theoretical expression of Eq. (2) evaluated numerically, considering that the excitation laser intensity spans more than four orders of magnitude. Since at the largest $\Omega$ in Fig. 3 the central peak continues to grow, some amount of light is at that point originating from other scatterers, perhaps neighboring detuned QDs. Nonetheless, when $\Omega/2\pi$=0.49 GHz, we can estimate that more than 90 \% of the detected light originates from scattering off the QD probed.

Photon statistics of the scattered light, shown in Fig. 4(a) below saturation (left panel), near saturation (middle panel) and above saturation (right panel), corroborate this via the observation of photon antibunching. Photon correlation measurements were performed with a Hanbury-Brown and Twiss setup \cite{michler2000qds} that uses a 2$\times$2 fiber coupler and two fiber coupled avalanche photon detectors. The detectors are nominally identical to those used in \cite{flagg2009} and thus an identical instrument response function was assumed.
We show the raw unprocessed normalized data as well as the theoretical normalized second order correlation function in the presence of spectral diffusion, $g_{SD}^{(2)}(t)$, convolved with the instrument response function (IRF) \cite{flagg2009}. It is the IRF, not stray background light, that causes the dip to rise above zero near $t$=0. $g_{SD}^{(2)}(t)$ is computed identically to Eq. (2), replacing $\tilde{g}(\nu,\omega,\omega_0')$ under the integral with $g^{(2)}(t,\omega,\omega_0')\bar{n}_{\infty}/\bar{n}_{\infty}(\omega=\omega_0)$, where $g^{(2)}(t,\omega,\omega_0')$ is the normalized the correlation function for the ideal two-level system. Above saturation Rabi oscillations, the time-domain analogue of the side bands of Fig. 2, are seen \cite{flagg2009}.

\begin{figure}[t!]
\includegraphics[width=3.1in]{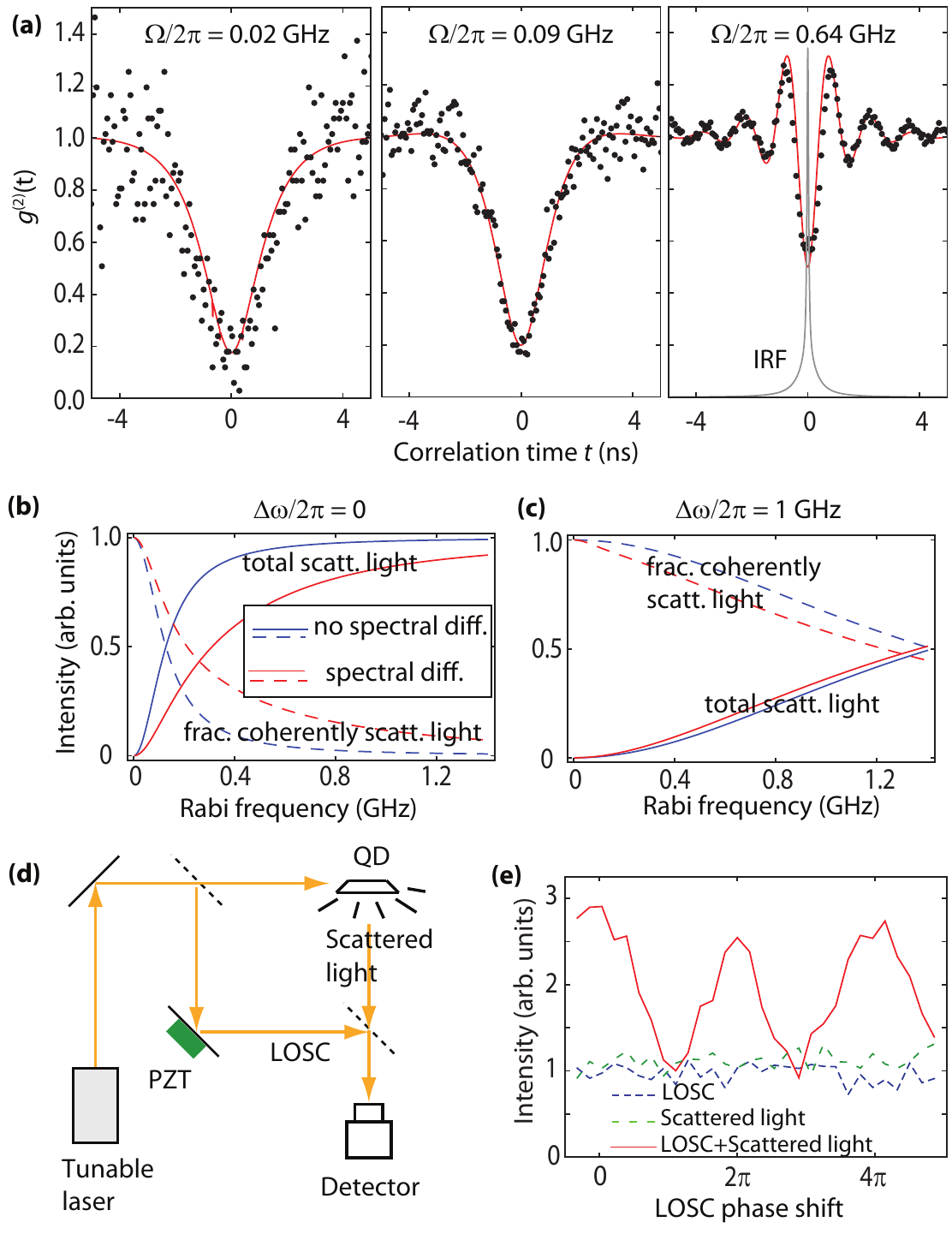}
\caption{\label{fig4} (Color online) (a) Photon statistics of scattered light for three different Rabi frequencies. When $\Omega>\kappa$, Rabi oscillations appear. (b) Plot of the total scattered light intensity (solid red trace) and the fraction of coherently scattered light intensity (dashed red trace), using the same parameters as those used in Fig. 3. For comparison the same are plotted in blue when not including spectral diffusion. (c) Same as in (b) but with a laser detuned by $\Delta\omega/2\pi$ = 1 GHz. (d) Measurement of mutual phase coherence between the coherently scattered light and a local oscillator (LOSC) by interferometry. (e) Intensity of light at the output of the beam splitter in (d) as a function of LOSC phase.}
\end{figure}

Spectral diffusion significantly alters the spectral and temporal characteristics of the scattering process compared to the ideal two-level system.
Figure 4(b) shows the total normalized scattered light intensity as a function of Rabi frequency for the parameters used in Fig. 3. using Eq. (2) (solid red trace) and Eq. (1) (solid blue trace). Correspondingly, the dashed lines of the same color represent the fraction of the intensity of coherently scattered light alone. The same are plotted in Fig. 4(c) but under significant laser detuning. A signature effect of the spectral diffusion is to cause an {\it increase} in the fraction of photons scattered coherently, as would any source of inhomogeneous broadening. At large values of $\Omega$, more light is actually scattered coherently with the laser off-resonance than with the laser at exact resonance. This is seen in Fig. 4(c) but is also directly visible in the bottom map of Fig. 2 where a minimum occurs along the line $\nu=\omega$ at resonance (inset).

Lastly we examine the phase coherence between the laser and the scattered light (below saturation) by combining the scattered light with a local oscillator (LOSC) signal at a beam splitter. We measure the fringe contrast obtained when varying the LOSC phase with a piezoelectric actuator (PZT) [Fig. 4(d)]. The fringe contrast we obtain here is $\approx$40\% as seen in Fig. 4(e) which can be understood by spectral diffusion that causes large fluctuations in photon flux at the beam splitter. In theory we expect spectral diffusion to reduce the fringe contrast (visibility) by a factor of order $\kappa/s\approx$0.25.

In summary, we have carried out high-resolution measurements of the light scattered coherently and incoherently by a single InAs QD, revealing in great detail how the scattering process evolves over more than four orders of magnitude of excitation laser intensity. The simple inclusion into Mollow's theory of spectral diffusion as a source of inhomogeneous broadening does faithfully reproduce both our observed spectra but also photon correlation and phase-coherence measurements. The insight that an apparent broadening rather than pure dephasing predominantly affects the scattering process has important implications for future use of QDs, for example in quantum repeaters \cite{flagg2010isp}. There we expect spectral diffusion to reduce the photon flux without however limiting two-photon indistinguishability \cite{ates2009}. The temporal flickering of the resonant scattering signal associated with the spectral diffusion may be reduced in future improved structures. For example it may be possible to control the timescale for this flickering by placing QDs in charge-tunable devices. Overall our work offers a complete picture of resonant light scattering relevant to a wide variety of solid-state nanostructures.

The authors acknowledge financial support from the National Science Foundation (NSF DMR-0906025 and CMMI-0928664) the National 
Natural Science Foundation of China (Grant No. 90921015).

\appendix
\section{Power spectrum in the presence of pure dephasing and detuning}

Using the same notation as in Ref. \cite{mollow1969psl}, we provide an expression for the power spectrum of the light scattered by a two-level system subject to an incident monochromatic laser, when including an additional off-diagonal decay (pure dephasing) rate $\gamma$. Unlike in Ref. \cite{muller2007rfc,flagg2009} we further assume that the detuning is not zero, i.e. $\Delta\omega\neq$ 0. 

The optical Bloch equations in the rotating-wave approximation read
\begin{equation}
  \frac{d}{dt} {\bf R}(t)={\bf M} \cdot {\bf R}(t)
  \label{stateSpaceForm1}
  \end{equation}
in which
\begin{equation}
\mathbf{M}= \left( \begin{array}{cccc}
  -\kappa & -i \Omega /2 & i \Omega /2 & 0 \\
  -i \Omega /2 & -\kappa /2-\gamma+i \Delta \omega& 0 & i \Omega /2 \\
  i \Omega /2 & 0 & -\kappa/2-\gamma-i \Delta \omega & -i \Omega /2 \\
  \kappa & i \Omega /2 & -i \Omega /2 & 0
  \label{Mmatrix}
\end{array} \right)
\end{equation}
and
\begin{equation}
\mathbf{R}(t)= \left( \begin{array}{c}
  n(t) \\
  \alpha (t) \\
  \alpha^{\ast} (t)\\
  m(t)
   \label{Rvector}
\end{array} \right)
\end{equation}
Here $ n(t)$=Tr$\{\rho(t) a^\dag a\} $, $ \alpha(t)$=Tr$\{\rho(t) a \}$,
$ \alpha^{\ast}(t)$=Tr$\{\rho(t) a^\dag \} $, and $ m(t)$=Tr$\{\rho(t) a
a^\dag\} $, where $a$, $a^\dag$, and $\rho$ are the lowering, raising, and density operators, respectively.
The diagonal and off-diagonal damping terms
$\frac{1}{T_1}=\kappa$ and $\frac{1}{T_2}=\gamma+\kappa/2$, respectively, have been
included using the usual master equation.
Equation \ref{stateSpaceForm1} has the steady state solution
\begin{equation}
  \alpha_\infty=\frac{i \Omega}{4} \frac{\kappa+2\gamma+2i\Delta\omega}
  {\Delta\omega^2+(\kappa+2\gamma)(\kappa+2\gamma+2\Omega^2/\kappa)/4}
  \label{stateSpaceForm2}
  \end{equation}
and
\begin{equation}
  n_\infty=\frac{\Omega^2}{4\kappa} \frac{\kappa+2\gamma}
  {\Delta\omega^2+(\kappa+2\gamma)(\kappa+2\gamma+2\Omega^2/\kappa)/4}
  \label{stateSpaceForm3}
  \end{equation}

Following Mollow \cite{mollow1969psl}, the power spectrum is obtained as the Fourier transform of the first-order correlation function, $g(\tau,t)\equiv \langle a^{\dag}(t)a(t+\tau) \rangle$, in steady-state ($t\rightarrow\infty$), as
\begin{equation}
  \tilde{g}(\nu)\equiv\int_{-\infty}^\infty g(\tau){e^{i\nu\tau}}d\tau=2 \mathrm{Re}\big(\hat{g}(-i\nu)\big)
\label{stateSpaceForm22}
   \end{equation}
where $\nu$ is the angular frequency of the scattered light, and $\hat{g}(s)$ is the Laplace transform of $g(\tau)$. In order to calculate $g(\tau)$ we make use of the quantum regression theorem \cite{scully} which states that for an
operator $\mathcal{O}$ whose expectation value is known to evolve
from time $t$ to time $t+\tau$ as
\begin{equation}
\langle \mathcal{O}(t+\tau)\rangle=\sum_j a_j(\tau) \langle
\mathcal{O}_j(t)\rangle
\end{equation}
the two-time correlation function $\langle\mathcal{O}_i(t)
\mathcal{O}(t+\tau) \mathcal{O}_k(t)\rangle$ can be calculated as a
function of single time expectation values as follows:
\begin{equation}
\langle \mathcal{O}_i(t) \mathcal{O}(t+\tau)
\mathcal{O}_k(t)\rangle=\sum_j a_j(\tau) \langle \mathcal{O}_i(t)
\mathcal{O}_j(t) \mathcal{O}_k(t)\rangle
\end{equation}
Given that the solution of equation \ref{stateSpaceForm1} can be written as
\begin{equation}
  {\bf R}(t+\tau)=e^{{\bf M}\tau}\cdot{\bf R}(t)
  \label{stateSpaceForm4}
  \end{equation}
we obtain
\begin{equation}
  g(\tau)=\big(e^{{\bf M}
  \tau}|_{2,2} n_\infty+e^{{\bf M}
  \tau}|_{2,4} \alpha^{\ast}_\infty\big)e^{-i\omega \tau-\delta_{FP}\tau/2}
 \label{stateSpaceForm6}
  \end{equation}
To account for the limited resolution, $\delta_{FP}$, of our scanning Fabry-Perot interferometer, we have multiplied the correlation function by $e^{-\delta_{FP}\tau/2}$ ($\tau\geq0$). Computing \ref{stateSpaceForm22} requires finding the Laplace transform of \ref{stateSpaceForm6}, which in turn involves computing the matrix $(\mathbb{I}s-{\bf M})^{-1}$, where $\mathbb{I}$ is the 4$\times$4 identity matrix. We finally obtain
\begin{displaymath}
  \tilde{g}(\nu)=2 \mathrm{Re}\big((\mathbb{I}(-i\nu+i\omega+\delta_{FP}/2)-{\bf M})^{-1}|_{2,2} n_\infty
  \end{displaymath}
\begin{equation}
+(\mathbb{I}(-i\nu+i\omega+\delta_{FP}/2)-{\bf M})^{-1}|_{2,4}\alpha^{\ast}_\infty\big)
\label{finalequation}
   \end{equation}
which for the special case $\gamma=0$ and $\delta_{FP}=0$ reduces to the expression
\begin{equation}
\tilde{g}(\nu)=2\pi|\alpha_{\infty}|^2\delta(\nu-\omega)
+\bar{n}_{\infty}\kappa\Omega^2\frac{(\nu-\omega)^2+\Omega^2/2+\kappa^2}{|f(i(\nu-\omega))|^2}
\end{equation}
given by Mollow [Eq. (1)], where $f(s)=s^3+2\kappa s^2+\big(\Omega^2+(\Delta\omega)^2+(5/4)\kappa^2 \big) s+\kappa\big(\frac{1}{2}\Omega^2+(\Delta\omega)^2+\frac{1}{4}\kappa^2 \big)$.

\end{document}